\documentclass[aps,prd,12point,twocolumn,nofootinbib,showpacs,superscriptaddress]{revtex4-1}
\def\theequation{\arabic{section}.\arabic{equation}}
\usepackage{psfrag}
\usepackage{subfigure}
\usepackage{color}
\usepackage{mathrsfs}
\usepackage{graphicx}
\usepackage{amssymb}
\usepackage{amsmath, amsthm}
\usepackage{epstopdf}
\usepackage{hyperref}
\usepackage{enumerate}

\newcommand{\be}{\begin{equation}}
\newcommand{\ee}{\end{equation}}
\newcommand{\bea}{\begin{eqnarray}}
\newcommand{\eea}{\end{eqnarray}}

\begin{document}
\def\theequation{\arabic{section}.\arabic{equation}} 

\title{Jordan frame no-hair for spherical scalar-tensor 
black holes}

\author{Valerio Faraoni}
\email[]{vfaraoni@ubishops.ca}
\affiliation{Physics Department, Bishop's University, 
2600 College Street, Sherbrooke, Qu\'ebec, Canada J1M~1Z7
}



\begin{abstract}
A no-hair theorem for spherical black holes in 
scalar-tensor gravity is presented. Contrary to 
the existing  theorems, which are proved in the 
Einstein conformal frame, this proof is performed entirely 
in the Jordan frame. The theorem is limited to spherical 
symmetry (instead of axisymmetry) but holds for 
non-constant Brans-Dicke couplings. 
\end{abstract}


\maketitle

\section{Introduction} 
\label{sec:1}
\setcounter{equation}{0}

There is little doubt that general relativity (GR) should 
eventually merge with quantum mechanics at the Planck 
scale, but it is not presently known how this should be 
done, although there are contenders to the role of a 
quantum gravity theory. It seems pretty clear, however, 
that quantizing gravity introduces extra degrees of freedom 
in addition to the metric tensor of GR, such as scalar 
and $p$-form fields, higher derivative terms in the field 
equations, and
other  deviations from GR. What is more, the 
acceleration of the present cosmic expansion discovered 
with type 
Ia 
supernovae is explained in the context of the GR-based 
$\Lambda$CDM model only by 
postulating a very exotic and completely {\em ad hoc} dark 
energy accounting for approximately 70\% of the energy 
content of the universe, the origin of which is a complete 
mystery (see Refs.~\cite{AmendolaTsujikawabook, 
LinderResletter} for reviews and references). A viable alternative to introducing dark 
energy is to modify gravity at cosmological scales 
\cite{CCT, CDTT, Vollick}, which has led to the  
popular class 
of $f(R)$ theories of gravity (see \cite{reviews} for 
reviews). To add to these motivations, GR is tested poorly 
in many regimes and assuming  its validity is currently 
more  a leap 
of faith than an experimental fact \cite{Bertietal2015, 
Psaltis}.

It is possible that some deviations from GR may be 
detectable at astrophysical  scales involving black holes 
and strong gravity. For this reason, many theoretical 
efforts have focused on the possibility of detecting black 
hole hair, 
which is forbidden in (vacuum) GR \cite{GRnohair, Hawkingprevious} but is 
allowed in other 
theories of gravity \cite{Thomasreviews, Herdeiroreview, 
Charmousisreview}. The Jebsen-Birkhoff 
theorem of GR 
\cite{Jebsen, Birkhoff}
states that the only vacuum, spherically symmetric, 
asymptotically flat solution of the Einstein equations with 
zero cosmological constant is the Schwarzschild solution. 
In addition to a host of no-hair theorems in GR 
\cite{GRnohair, Hawkingprevious}, similar theorems have been proved also in 
certain other 
theories of gravity, including the theory  of 
a scalar field non-minimally coupled to gravity, 
Brans-Dicke and scalar-tensor theories \cite{Johnson, 
Bekensteinreview,  
HawkingBD, myBirkhoff, SotiriouFaraoniPRL, Romano}, and 
Horndeski 
gravity (see \cite{Thomasreviews,  
Herdeiroreview,Charmousisreview} for recent 
reviews). Here we focus on scalar-tensor theories of 
gravity, which include $f(R)$ gravity as a subclass 
\cite{reviews}. A celebrated no-hair theorem 
by Hawking \cite{HawkingBD} states 
that all asymptotically flat, axisymmetric, stationary 
black holes in Brans-Dicke theory with constant 
Brans-Dicke coupling $\omega$ and zero potential 
for the Brans-Dicke gravitational scalar field $\phi$ are 
GR ({\em i.e.}, Kerr or Kerr-Newman) black 
holes. This result is important because these black 
holes represent the endpoint of gravitational collapse 
(indeed, the endpoints of collapse calculated numerically 
in Brans-Dicke gravity are invariably GR black holes 
\cite{BDcollapse}). 
 This theorem has  been generalized recently 
to include more general scalar-tensor theories  in which 
the Brans-Dicke coupling becomes a function 
$\omega(\phi)$  of the scalar and 
there is a potential $V(\phi)$ 
\cite{SotiriouFaraoniPRL}. The more recent 
Ref.~\cite{Romano} 
includes the case in which the black hole has de Sitter 
(instead of Minkowski) asymptotics.

There are, however, important limitations in these results: 
first, black holes can be perturbed from their stationary 
equilibrium state by surrounding matter, which can take the 
form of a 
star or of another black hole in a binary system, or 
of matter flowing in  
from a surrounding accretion disk. The resulting 
dynamics may give away some deviations from GR, but 
the no-hair theorems fail to catch this physics. Second, 
the assumption of asymptotic flatness is, at best, an 
approximation which may be good in practice, but 
ultimately realistic black 
holes are embedded in the universe. This universe does 
not have  a de 
Sitter asymptotics, but a 
Friedmann-Lema\^itre-Robertson-Walker (FLRW) one. 
Therefore, 
although much effort has gone into establishing more 
general no-hair theorems or 
probing the scalar hair deviations from GR, there are still 
important gaps in our understanding of these issues.

Independent motivation for extending the available no-hair 
theorems of scalar-tensor gravity comes from the fact that 
all these theorems thus far have been 
obtained in 
the Einstein conformal frame representation of the  
theory. While it is reasonable to regard the Jordan and 
the Einstein conformal frames as equivalent 
representations, as proposed long ago by Dicke 
\cite{Dicke}, there 
has been much discussion as to whether these two frames are 
physically equivalent \cite{debate}, 
and the debate has not 
settled. Part of the problem consists of defining what 
``physically 
equivalent'' precisely means. We do not weigh in on 
this issue here but, for the skeptic, we provide a  no-hair 
theorem which is entirely based in the Jordan frame. It  
extends Hawking's no-hair theorems by allowing for 
non-constant Brans-Dicke couplings, but it is 
restricted to {\em spherical} black holes, contrary to 
Hawking's result and its more recent generalizations 
\cite{SotiriouFaraoniPRL, Romano} which hold for more 
realistic {\em axiysimmetric} ({\em i.e.}, rotating) black 
holes. Therefore, the  value of our theorem lies not so much 
in extending previous results (spherical 
symmetry is indeed a serious restriction), but rather in 
showing 
that the Jordan frame allows for essentially the same 
results that can be obtained in the Einstein frame, 
although the techniques used are quite different (we reason 
on the local field equations instead of using integrals of 
these equations over the horizon and at spatial infinity).

We begin by reviewing the essential features of 
scalar-tensor gravity, in  
the notation of~\cite{Waldbook}. Scalar-tensor theories 
are described by the action \cite{BransDicke, ST}
\bea 
S_{ST} &=&\frac{1}{16\pi}\int d^4 x \sqrt{-g}\left[\phi 
R-\frac{\omega(\phi)}{\phi} \, 
g^{ab}\nabla_a\phi\nabla_b\phi \right. \nonumber\\
&&\nonumber\\
&\, & \left. -V(\phi)\right]+S^{(m)} 
\label{STaction}
\eea
in the Jordan conformal frame, where $S^{(m)}$ is the 
matter part of the action, $R$ is the Ricci scalar, $\phi$ 
is the Brans-Dicke-like scalar field, the 
function $\omega(\phi)$ 
is the Brans-Dicke coupling, and $V(\phi)$ is a potential. 
The 
field equations obtained by varying~(\ref{STaction})  can 
be 
written in the form of effective Einstein equations as
\bea
R_{ab}-\frac{1}{2}g_{ab}R & = 
&\frac{8\pi}{\phi}T_{ab}^{(m)} \nonumber\\
&&\nonumber\\
&\, & + 
\frac{\omega(\phi)}{\phi^2}\left(\nabla_a\phi\nabla_b\phi 
-\frac{1}{2}g_{ab} \, \nabla^c\phi\nabla_c\phi\right) 
\nonumber\\
&&\nonumber\\
& \, & + \frac{1}{\phi}\left(\nabla_a\nabla_b\phi 
-g_{ab}\Box\phi\right)-\frac{V(\phi)}{2\phi} 
g_{ab} \nonumber\\
&&\nonumber\\
& \equiv & \frac{8\pi}{\phi}\left( 
T_{ab}^{(m)}+T_{ab}^{(\phi)}\right) \,,\label{STfield1}\\
&&\nonumber\\
\left( 2\omega+3 \right) \Box\phi & = & 8\pi 
T^{(m)}-\frac{d\omega}{d\phi} 
\, \nabla^c\phi\nabla_c\phi+\phi \, \frac{dV}{d\phi}-2V  
\,,\nonumber\\
&& \label{fephi}
\eea 
where $T_{ab}^{(m)}$ is the matter energy-momentum tensor 
and we assume that $\phi>0$ in conjunction with 
$\omega>-3/2$ to guarantee the positivity of the effective 
gravitational coupling \cite{Nordtvedt}
\be
G_{eff}=\frac{2\left( \omega+2 \right)}{\left( 2\omega 
+3 \right) \phi} \,.
\ee
By imposing that the matter 
energy-momentum tensor $T_{ab}^{(m)}$ vanishes, a 
time-dependent $ T_{ab}^{(\phi)}$ would spoil the validity 
of the Jebsen-Birkhoff 
theorem and only the assumption that $\phi$ is 
time-independent (or that $T_{ab}^{(\phi)}$ is static, as 
for a time-dependent stealth field $\phi$ \cite{stealth} 
or for other solutions which do not enjoy the symmetries of 
the metric \cite{Thomasreviews}),   
restores the staticity of spherical geometries.

When $\phi=$~const.$ \equiv \phi_0$, 
Eq.~(\ref{STfield1}) reduces to
\be
R_{ab}-\frac{1}{2}\, g_{ab}R =\frac{8\pi }{\phi_0}\, 
T^{(m)}_{ab}-\frac{V_0}{2\phi_0}\, 
g_{ab} \,,
\ee
where $V_0 \equiv V(\phi_0)$ if a scalar field potential is 
present, hence the theory degenerates into GR with the 
cosmological constant $\Lambda \equiv V_0/(2\phi_0)$.

By performing the conformal transformation of the metric 
and redefining non-linearly the Brans-Dicke-like scalar as 
in
\begin{eqnarray}
&& g_{ab}\rightarrow \tilde{g}_{ab}=\Omega^2 \, g_{ab}, 
\,\,\,\,\,\,\, \Omega=\sqrt{\phi} \,, \label{confo1}\\
&&\nonumber \\
&& d\tilde{\phi}=\sqrt{ \frac{ 2\omega(\phi)+3 }{16\pi}} \, 
\frac{d\phi}{\phi} \,, \label{confo2}
\end{eqnarray}
the scalar-tensor action~(\ref{STaction}) 
assumes the Einstein frame form
\begin{eqnarray}
 S_{ST} &=&  \int d^4x \, \sqrt{-\tilde{g}} \left[
\frac{\tilde{R}}{16\pi} -\frac{1}{2}\, \tilde{g}^{ab} 
\tilde{\nabla}_a \tilde{\phi} 
\tilde{\nabla}_b \tilde{\phi} 
-U(\tilde{\phi})  \right. \nonumber\\
&&\nonumber\\
&\, & \left. +     \frac{ {\cal 
L}^{(m)} }{ \phi^2} \right]  \,,
\end{eqnarray}
where a tilde denotes quantities associated with the 
rescaled metric $\tilde{g}_{ab}$, ${\cal L}^{(m)}$ is the 
matter Lagrangian density, and
\be \label{questaA}
U( \tilde{\phi})=\frac{V\left[ \phi( \tilde{\phi}) 
\right]}{\left[ \phi ( \tilde{\phi}) \right]^2} \,.
\ee 
This is formally the action of GR with 
a minimally coupled scalar field $\tilde{\phi}$ but with 
the important difference that this scalar now couples 
explicitly to matter  \cite{Dicke}. The Einstein frame 
field equations are
\begin{eqnarray}
&&  \tilde{R}_{ab}-\frac{1}{2}\, 
\tilde{g}_{ab}\tilde{R}=\frac{8\pi}{\phi^2}\, T_{ab}^{(m)} +
8\pi \tilde{T}_{ab}^{(\tilde{\phi})} \,,\\
&&\nonumber\\
&& \tilde{\Box} \tilde{\phi} -\frac{ 
dU}{ d\tilde{\phi} }=\frac{8\pi T^{(m)} }{\phi^2} 
\,,\label{Eframeeqforscalar}
\end{eqnarray}
where 
\be
\tilde{ T}_{ab}^{(\tilde{\phi})}=\tilde{\nabla}_a \tilde{\phi}
\tilde{\nabla}_b \tilde{\phi} - \frac{1}{2}\, \tilde{g}_{ab} \,
\tilde{g}^{cd}\tilde{\nabla}_c \tilde{\phi}
\tilde{\nabla}_d \tilde{\phi} -\frac{U( \tilde{\phi})}{2}
\, \, \tilde{g}_{ab}  
\ee
is the canonical stress-energy tensor for a scalar field 
minimally coupled with the curvature, which satisfies the 
weak energy condition if $V\geq 0$. 

When the scalar field $\tilde{\phi}$ 
is constant (which only happens if its Jordan frame cousin 
$\phi$ is constant) then one obtains the same equations of 
motion as in GR (with a cosmological 
constant if $U( \tilde{\phi}) \neq 0$, which is equivalent 
to $V(\phi) \neq 0$, as follows from 
Eq.~(\ref{questaA})).

We are now ready to prove the new no-hair result for 
scalar-tensor black holes.

\section{Jordan frame no-hair}
\label{sec:2}
\setcounter{equation}{0}

Consider a scalar-tensor theory 
in which the Brans-Dicke coupling is a function $\omega ( 
\phi)$ \cite{ST} with  
a potential $V(\phi)$ and  assume  
spherical symmetry. Up to singular exceptions, the most 
general spherically symmetric 
line element can be 
written as
\be\label{spher}
ds^2=-A^2(t, r)dt^2+B^2(t, r)dr^2+r^2 \left( d\theta^2+\sin^2 
\theta \, d\varphi^2 \right) 
\ee
in spherical coordinates $\left( t,r, \theta, \varphi 
\right)$. We assume also that the Brans-Dicke scalar  
is spherically symmetric, $\phi=\phi(t,r)$. The only  
non-vanishing  Christoffel symbols  are 
\begin{eqnarray}
&&  \Gamma^0_{00}=\frac{\dot{A}}{A} \,, \;\;\;\;\;\;
\Gamma^0_{01}=\Gamma^0_{10} = \frac{A'}{A} \,, \;\;\;\;\;\;
\Gamma^0_{11}=\frac{B\dot{B}}{A^2} \,,  \nonumber\\
&&\\
&&  \Gamma^1_{00}=\frac{AA'}{B^2} \,, \;\;\;\;\;\;
\Gamma^1_{01}=\Gamma^1_{10} = \frac{\dot{B}}{B} \,, \;\;\;\;\;\;
\Gamma^1_{11}=\frac{B'}{B} \,, \nonumber\\
&&\\
&& \Gamma^1_{22}=- \frac{r}{B^2} \,, \;\;\;\;\;\;
\Gamma^1_{33}=- \frac{r}{B^2}\, \sin^2 \theta \,, \\
&& \nonumber\\
&& \Gamma^2_{12}=\Gamma^2_{21}=\frac{1}{r} \,, \;\;\;\;\;\;
\Gamma^2_{33}=-\cos \theta  \,, \\
&& \nonumber\\
&& \Gamma^3_{13}=\Gamma^3_{31}= \frac{1}{r}   \,, \;\;\;\;\;\;
\Gamma^3_{23}=\Gamma^3_{32}= \cot\theta  \, .
\label{Christoffel}
\end{eqnarray}
Using these quantities and the notation  $ {}^{\centerdot} 
\equiv \partial_t $ and 
$' \equiv  \partial_r$  we have, in the 
geometry~(\ref{spher}),
\bea
\nabla^c\phi\nabla_c\phi  &=&  -\frac{\dot{\phi}^2}{A^2} 
+\frac{\phi'^2}{B^2} \,,\\
&&\nonumber\\
\Box\phi &=& -\frac{1}{A^2}\left(\ddot{\phi} 
-\frac{\dot{A}}{A} \,  \dot{\phi}-\frac{AA'}{B^2} \, 
\phi'\right) \nonumber\\
&&\nonumber\\
&\, &  +\frac{1}{B^2}\left(\phi''-\frac{B\dot{B}}{A^2} \, 
\dot{\phi} 
-\frac{B'}{B} \, \phi'\right) \nonumber\\
&&\nonumber\\
&\, & +\frac{2\phi'}{r B^2} \,. \label{boxphi} 
\eea
Since the endpoint of gravitational collapse must be a 
stationary black hole (cf. \cite{HawkingBD, 
SotiriouFaraoniPRL, Romano}), we assume a  
static scalar field and metric ($\dot{\phi}=0, 
\dot{A}=\dot{B}=0$). In electrovacuo or in the presence of 
conformally invariant matter\footnote{This 
situation includes the Maxwell field, a conformally 
coupled scalar field with zero or quartic potential, 
and a radiation fluid.}  $T^{(m)}=0$ and  
the d'Alembertian~(\ref{boxphi}) reduces to 
\be
\Box\phi=\frac{1}{B^2}\left[\phi''+\left(\frac{A'}{A} 
-\frac{B'}{B} +\frac{2}{r} \right) \phi'\right]
\ee
while Eq.~(\ref{fephi}) becomes
\bea 
(2\omega+3)\left[\phi''+\left(\frac{A'}{A} 
-\frac{B'}{B} +\frac{2}{r} \right)\phi'\right] \nonumber\\
\nonumber\\
=  - \, \frac{d\omega}{d\phi} \, \phi'^2  +B^2\left( \phi 
V_{\phi}-2V \right) \,, \label{boh}
\eea
where $V_{\phi} \equiv dV/d\phi$. 
Since the line element~(\ref{spher}) is spherically 
symmetric and is expressed in Schwarzschild-like 
coordinates employing the areal radius, the apparent 
horizon is the locus $\nabla_cr\nabla^c r=0$ 
\cite{MisnerSharp, Boothreview, NielsenGRGreview, mylastbook} and it 
coincides with the event horizon because the 
geometry is static \cite{HawkingEllis}. The horizon radius 
$r_H$ is given by the (positive) root of 
$g^{rr}=0$ in these coordinates. The horizon at $g^{11}=0$ 
is obtained as $B\rightarrow\infty$. To proceed, note that 
\be
\nabla^c\phi\nabla_c\phi =\left(\phi'\right)^2 
\nabla^cr\nabla_cr  =\frac{\left(\phi'\right)^2}{B^2}=0   
\ee
on the horizon, provided that $\phi$ and $\phi'$ are well 
defined there.  

Let us assume now, as in the original Brans-Dicke theory 
\cite{BransDicke} and in Hawking's theorem \cite{HawkingBD} 
that 
the potential $V(\phi)$ vanishes identically.\footnote{The 
same result is obtained if $V(\phi)=m^2 \phi^2/2$, in which 
case the potential disappears from the 
Klein-Gordon-like Eq.~(\ref{fephi}) ruling the 
dynamics of $\phi$ \cite{Santos00}, but not from the field 
equations~(\ref{STfield1}).}  This is clearly a limitation 
but it allows one to proceed with the proof. 
Equation~(\ref{boh}) then gives
\bea
\frac{\phi''}{\phi'}+\frac{\omega_{\phi} \phi'}{2\omega+3} 
+\frac{A'}{A} -\frac{B'}{B} +\frac{2}{r}  \nonumber\\ 
\nonumber\\
= \left[\ln\left(  \frac{Ar^2 \phi'}{B}  
\, \sqrt{2\omega+3}\right)\right]' = 0 \,,
\eea
which integrates to
\be
\frac{\phi'(r)}{B}=\frac{C_0}{\sqrt{2\omega+3} 
\, Ar^2} \,,
\ee
with $C_0$ being an integration constant.  Assuming that $\phi$ 
and $ \phi'$ are finite on the 
black hole horizon, the condition $1/B^2 \rightarrow 0$ as  
$ r\rightarrow r_H $  yields 
\be
\frac{C_0}{r_H^2 A(r_H)}=0
\ee
and the constant $C_0 $ must vanish. This means that 
\be
\phi'(r)=0\;\;\;\;\;\;\forall\;r\geq r_H 
\ee
and $\phi=$~constant ~$\forall\;r\geq r_H $. Therefore, the 
black hole must reduce to a GR black hole.   As a 
conclusion, in the scalar-tensor theory~(\ref{STaction}) 
and under the assumptions above, a 
black hole is necessarily a GR black hole. 
 This proof shows 
that the limitations imposed in the usual theorems are not 
due to the choice of conformal frame and that it is worth 
attempting to extend the existing no-hair theorems using 
the Jordan frame.

\section{Conclusions}
\label{sec:4}
\setcounter{equation}{0}

The proof of the previous section is completed entirely in 
the  Jordan frame. 
Although the techniques used are different than the usual 
Einstein frame techniques, the result shows that there is 
no {\em a priori} reason why the Jordan frame should not 
be useful for this kind of study. The fact that the  
Jordan frame proof presented cannot be completed when a 
non-trivial scalar field potential is present is due to 
purely technical reasons (the fact that it becomes 
impossible to integrate analytically the non-linear 
ordinary differential equation~({\ref{boh})), and not to 
reasons of principle. 
The Jordan frame proof does not require the weak energy 
condition to be satisfied by the scalar field (in general 
it is not satisfied, contrary to the Einstein 
frame), and the proof is restricted to spherically 
symmetric black holes.

We assumed staticity of the metric and 
scalar field as features characterizing the endpoint of 
collapse.\footnote{Staticity of the 
metric and spherical symmetry do not automatically imply 
that the scalar field is static, as demonstrated by various  
scalar field solutions of scalar-tensor gravity 
\cite{Herdeiroreview, Thomasreviews, stealth}.} Strictly 
speaking, we did not assume asymptotic flatness or de 
Sitter asymptotics. Removing 
the assumption of asymptotic flatness present in most 
no-hair theorems (or that of de Sitter asymptotics)  and 
allowing for 
more general asymptotics (for example FLRW), would be a 
major step forward in the existing no-hair theorems. 
However, we do not claim such a generalization here because 
1)~it is hard to see how physically significant 
non-asymptotically flat black holes 
could arise {\em in vacuo} in the absence of a scalar field 
potential 
$V(\phi)$ and~2) when the field $\phi$ becomes constant, 
which 
happens in both our proof and those of the 
existing no-hair theorems, the 
theory reduces to GR but 
the only possible asymptotics are then constrained to be 
Minkowski or de Sitter. More 
interesting situations, for example those involving $f(R)$ 
black holes, must allow for FLRW asymptotics. As remarked 
in \cite{SotiriouFaraoniPRL, Thomasreviews}, it is true 
that one then expects variability on cosmological time 
scales much larger than the local (astrophysical) time 
scales, and that these variations could be neglected for 
the purposes of astrophysics, but one would nevertheless 
like to have a more complete 
picture, which is not available at the moment.

\begin{acknowledgments}

This work is supported by the Natural Sciences and 
Engineering Research Council of Canada. 

\end{acknowledgments}


\end{document}